\documentclass[11pt,a4paper]{article}
\usepackage{amsfonts}

\def\be{\begin{equation}}
\def\ee{\end{equation}}
\def\bea{\begin{eqnarray}}
\def\eea{\end{eqnarray}}
\def\Gammabol{{\stackrel{\circ}{\Gamma}}{}}
\def\Abol{{\stackrel{~\circ}{A}}{}}

\def\Rbol{{\stackrel{\circ}{R}}{}}

\def\Gammabol{{\stackrel{\circ}{\Gamma}}{}}

\def\Dbol{{\stackrel{\circ}{\mathcal D}}{}}
\def\nabol{{\stackrel{\circ}{\nabla}}{}}
\def\Lbol{{\stackrel{\circ}{\mathcal L}}{}}

\def\Gammaw{{\stackrel{\bullet}{\Gamma}}{}}

\def\Aw{{\stackrel{~\bullet}{A}}{}}

\def\tw{{\stackrel{\bullet}{t}}{}}
\def\L{{\mathcal L}{}}
\def\Lw{{\stackrel{\bullet}{\mathcal L}}{}}
\def\Tw{{\stackrel{\bullet}{T}}{}}
\def\Kw{{\stackrel{\bullet}{K}}{}}
\def\nablaw{{\stackrel{\bullet}{\nabla}}{}}
\def\Dw{{\stackrel{\bullet}{\mathcal D}}{}}
\def\dw{{\stackrel{\bullet}{D}}{}}
\def\Sw{{\stackrel{\bullet}{S}}{}}
\def\onehalf{{\textstyle{\frac{1}{2}}}}

\begin{document}
\noindent
{\Large \bf Gravitation: in search of the missing torsion}
\vskip 0.7cm
\noindent
{\bf R. Aldrovandi and J. G. Pereira} \\
{\it Instituto de F\'{\i}sica Te\'orica},
{\it Universidade Estadual Paulista} \\
{\it Rua Pamplona 145},
{\it 01405-900 S\~ao Paulo, Brazil}

\vskip 0.8cm
\noindent
{\bf Abstract~} A linear Lorentz connection has always two fundamental derived characteristics: curvature and torsion. The latter is assumed to vanish in general relativity. Three gravitational models involving non-vanishing torsion are examined: teleparallel gravity, Einstein--Cartan, and new general relativity. Their dependability is critically examined. Although a final answer can only be given by experience, it is argued that teleparallel gravity provides the most consistent approach.


\section{Introduction}

Gravitation has a quite peculiar property: particles with different masses and different compositions feel it in such a way that all of them acquire the same acceleration and, given the same initial conditions, follow the same path. Such universality of response --- usually referred to as {\it universality of free fall} --- is the most fundamental characteristic of the gravitational interaction \cite{mtw}. It is unique, peculiar to gravitation: no other basic interaction of nature has it. Universality of free fall is usually identified with the weak equivalence principle, which establishes the equality between inertial and gravitational masses. In fact, in order to move along the same trajectory, the motion of different particles must be independent of their masses, which have then to be canceled out from the equations of motion. Since this cancellation can only be made when the inertial and gravitational masses coincide, this coincidence will naturally imply universality.

General relativity, Einstein's theory for gravitation, is fundamentally based on the weak equivalence principle. In fact, to comply with universality, the presence of a gravitational field is supposed to produce {\em curvature in spacetime}, the gravitational interaction being achieved by letting (spinless) particles to follow the {\it geodesics} of the curved spacetime. In other words, the connection describing the gravitational interaction is assumed to have non-vanishing curvature, but vanishing torsion. 

And now comes the important point: a general spacetime--rooted (or Lorentz) connection has {\em two} fundamental properties: curvature {\em and}\, torsion. Why should then matter produce {\it only curvature}? Was Einstein wrong when he made this assumption? Does torsion play any role in gravitation? The purpose of these notes is to discuss possible answers to this question, as well as to analyze their  theoretical and experimental consistencies \cite{moriond}.

In order to do that we begin by introducing, in the next section, some fundamental concepts related to spacetime and gravitation. Then, in section 3, we briefly review the main points of general relativity, a theory in which torsion is assumed to vanish from the very beginning. Section 4 gives a discussion of teleparallel gravity, a theory in which, instead of torsion, curvature is assumed to vanish. In spite of this fundamental difference, general relativity and teleparallel gravity are found to provide completely equivalent descriptions of the gravitational interaction. According to these theories, therefore, curvature and torsion are equivalent ways of describing gravitation, and consequently related to the same gravitational degrees of freedom. Section 5 briefly outlines Einstein--Cartan theory, which presupposes the simultaneous existence of curvature and torsion. According to it, torsion becomes important only in the presence of intrinsic spin, and new physical phenomena --- ignored by general relativity --- are predicted to exist in its presence. For spinless matter, it coincides with general relativity. In this theory, therefore, curvature and torsion are related to different degrees of freedom of gravity. Section 6 discusses new general relativity,  a generalized teleparallel model with three free--parameters, which should be determined by experience. Differently from Einstein--Cartan, in this theory torsion is not necessarily related to intrinsic spin. However, similarly to Einstein--Cartan, torsion is assumed to be responsible for describing possible corrections to general relativity --- and consequently to the teleparallell equivalent of general relativity. Also in this case, therefore, curvature and torsion are related to different degrees of freedom of gravity. Finally, section 7 is devoted to a comparative discussion of the different interpretations for torsion.

\section{Basic concepts}

\subsection{Linear frames and tetrads}
\label{subsecFrames}

The geometrical setting of any theory for gravitation is the tangent bundle, a natural construction always present on spacetime \cite{livro}.   At each point of spacetime --- the bundle base space --- there is always a tangent space  --- the fiber --- on which a spacetime transformation group acts. In the case of the Lorentz group, the tangent space provides a representation for the group, the vector representation. The  bundle formalism provides a natural extension to other representations --- tensorial or spinorial. In what follows, we are going to use the Greek alphabet $(\mu, \nu, \rho, \dots = 0,1,2,3)$ to denote indices related to spacetime, and the first half of the Latin alphabet $(a,b,c, \dots = 0,1,2,3)$ to denote indices related to the tangent space, assumed to be a Minkowski spacetime with metric
\be
\eta_{ab} = \mbox{diag}(+1,-1,-1,-1).
\ee
The second half of the Latin alphabet $(i,j,k, \dots = 1,2,3)$ will be reserved for space
indices. Spacetime coordinates, therefore, will be denoted by $\{x^\mu\}$, whereas the
tangent space coordinates will be denoted by $\{x^a\}$. Such coordinate systems define, on
their domains of definition, local bases for vector fields, formed by the sets of
gradients
\be
\{\partial_\mu\} \equiv \{{\partial}/{\partial x^\mu}\} \quad \mbox{and} \quad
\{\partial_a\} \equiv \{{\partial}/{\partial x^a}\},
\ee
as well as bases $\{dx^\mu\}$ and
$\{dx^a\}$ for covector fields, or differentials. These bases are dual, in
the sense that
\be
dx^\mu \, ({\partial_\nu}) = \delta^\mu_\nu \quad \mbox{and} \quad
dx^a \, ({\partial_b}) = \delta^a_b.
\ee
On the respective domains of definition, any vector or covector field 
can be expressed in terms of these bases, which can furthermore be extended by
direct product to constitute bases for general tensor fields.

A {\em holonomic} (or coordinate) base like $\{{\partial_a}\}$, related to coordinates, is a very particular case of linear base. Any set of four linearly independent fields
$\{e_{a}\}$  will form another base, and will have a dual
$\{e^{a}\}$ whose members are such that $e^{a}(e_b) =
\delta^a_b$. These frame fields are the general linear bases on the spacetime
differentiable manifold, whose set, under conditions making of it also a
differentiable manifold, constitutes the bundle of linear frames.  Of
course, on the common domains they are defined, the members of a base can be
written in terms of the members of the other, that is,
\be
e_a = e_a{}^\mu \, \partial_\mu, \quad e^{a} = e^{a}{}_\mu \, dx^\mu,
\ee
and conversely. These frames, with their bundles, are constitutive parts of
spacetime. They are automatically present as soon as spacetime is taken to be a
differentiable manifold \cite{livro}.

We are going to use the notation $\{h_{a}, h^{a}\}$ for a generic tetrad field (or simply ``tetrad''), a field of  linear frames connected with the presence of a gravitational field. Consider the spacetime metric $g$ with components $g_{\mu \nu}$, in some dual holonomic base $\{d x^{\mu}\}$:
\begin{equation}
g = g_{\mu \nu} dx^{\mu} \otimes dx^{\nu} = g_{\mu \nu} dx^{\mu} dx^{\nu}.
\label{eq:Riemetric}
\end{equation}
A tetrad field $h_{a} = h_{a}{}^{\mu} \, {\partial_{\mu}}$ will relate $g$ to the tangent--space metric $\eta = \eta_{a b} \, dx^a dx^b$ by
\begin{equation}
\eta_{a b} = g(h_{a},h_{b}) = g_{\mu \nu} \, h_{a}{}^{\mu} h_{b}{}^{\nu}.
\label{eq:gtoeta}
\end{equation}
This means that a tetrad field is a linear frame whose members $h_{a}$ are
(pseudo) orthogonal by the metric $g$. The components of the dual base members
$h^{a} = h^{a}{}_{\nu} dx^{\nu}$ satisfy
\begin{equation}
h^{a}{}_{\mu} h_{a}{}^{\nu} = \delta_{\mu}^{\nu} \quad {\rm and} \quad
h^{a}{}_{\mu} h_{b}{}^{\mu} = \delta^{a}_{b},
\label{eq:tetradprops1}
\end{equation}
so that Eq.~(\ref{eq:gtoeta}) has the converse
\begin{equation}
g_{\mu \nu} = \eta_{a b} \, h^{a}{}_{\mu} h^{b}{}_{\nu}.
\label{eq:tettomet}
\end{equation}

Anholonomy --- the property of a differential form which is not the differential of anything, or of a vector field which is not a gradient --- is commonplace in
many chapters of Physics. Heat and work, for instance, are typical anholonomic coordinates on the space of thermodynamic variables, and the angular velocity of a generic rigid
body is a classical example of anholonomic velocity. In the context of gravitation,
anholonomy is related, through the equivalence principle, to the very existence of a
gravitational field \cite{ABP03}. Given a Riemannian metric as in~(\ref{eq:tettomet}), the presence or absence of a gravitational field is fixed by the anholonomic or holonomic character of the
forms $h^{a} = h^{a}{}_{\nu} dx^{\nu}$. We can think of a coordinate change $\{x^a\}
\leftrightarrow \{x^\mu\}$ represented by
\be
dx^a = \left(\partial_{\mu} x^a\right) \, dx^\mu \quad \mathrm{and} \quad dx^\mu =
\left(\partial_a x^{\mu}\right) \, dx^a. 
\ee
The 1-form $dx^a$ is holonomic, just the differential of the coordinate
$x^a$, and the objects $\partial_{\mu} x^a$ are
the components of the holonomic form $dx^a$ written in the base $\{dx^\mu\}$, with
$\partial_a x^{\mu}$ its inverse. Thus, such a coordinate change is just a change of
holonomic bases of 1-forms. For the dual base we have the relations
\be
\partial_\mu = \left(\partial_{\mu} x^a\right) \,
\partial_a \quad \mathrm{and} \quad \partial_a =
\left(\partial_a x^{\mu}\right) \, \partial_\mu.
\label{1}
\ee

Take now a dual base $h^a$ such that $d h^a \ne 0$, that is, not formed by differentials.
Apply the anholonomic 1-forms $h^a$ to ${\partial}/{\partial_\mu}$. The result, $h^a{}_\mu
= h^a \, ({\partial_\mu})$, is the component of  $h^a$ = $h^a{}_\mu dx^\mu$ along
$dx^\mu$. The procedure can be inverted when the $h^a$'s are linearly independent, and
defines vector fields $h_a$ = $h_a{}^\mu {\partial_\mu}$ which are not gradients. Because
closed forms are locally exact, holonomy/anholonomy can be given a trivial criterion: a
form is holonomic {\it iff} its exterior derivative vanishes.  A holonomic tetrad will
always be of the form $h^{a} = dx^a$ for some coordinate  set $\{x^a\}$. For such a tetrad,
 tensor (\ref{eq:tettomet}) would simply give  the components of the Lorentz metric
$\eta$ in the coordinate system $\{x^\mu\}$.

An anholonomic basis $\{h_{a}\}$ satisfies the commutation table
\begin{equation}
[h_{a}, h_{b}] = f^{c}{}_{a b}\ h_{c},
\label{eq:comtable}
\end{equation}
with $f^{c}{}_{a b}$ the so-called structure coefficients, or coefficients  of anholonomy. 
The frame $\{{\partial_{\mu}}\}$ has been presented above as holonomic
precisely because its members commute with each other. The dual expression of the
commutation table above is the Cartan structure equation
\begin{equation}
d h^{c} = - \onehalf \, f^{c}{}_{a b}\ h^{a} \wedge h^{b} = \onehalf \, 
(\partial_\mu h^c{}_\nu - \partial_\nu h^c{}_\mu)\ dx^\mu \wedge dx^\nu.
\label{eq:dualcomtable}
\end{equation}
The structure coefficients represent the curls of the base members:
\begin{equation}
f^c{}_{a b}  = h^c ([h_a, h_b]) = h_a{}^{\mu} h_b{}^{\nu} (\partial_\nu
h^c{}_{\mu} - 
\partial_\mu h^c{}_{\nu} ) =  h^c{}_{\mu} [h_a(h_b{}^{\mu}) - 
h_b(h_a{}^{\mu})].
\label{fcab}
\end{equation}
If $f^{c}{}_{a b}$ = $0$, then $d h^{a} = 0$ implies the local existence of functions
(coordinates) $x^a$ such that $h^{a}$ = $dx^a$. Nothing really new: the tetrads are gradients only when their curls vanish. 

\subsection{Connections}

In order to define derivatives with a well-defined tensor behavior (that is, which are
covariant), it is essential to introduce connections $\Gamma^\lambda{}_{\mu \nu}$, which
are vectors in the last index but whose non--tensorial behavior in the first two indices
compensates the non--tensoriality of the ordinary derivatives. Linear connections have a
great degree of intimacy with spacetime because they are defined on the bundle of linear
frames, which is a constitutive part of its manifold structure. That bundle has some
properties not found in the bundles related to {\it internal} gauge theories. 
Mainly, it exhibits soldering, which leads to the existence of torsion for every
connection \cite{koba}. Linear connections --- in particular, Lorentz connections --- always
have torsion, while internal gauge potentials have not. 

It is important to remark that, from a formal point of view, curvature and
torsion are properties of a connection~\cite{koba}. Strictly speaking, there are no such things as curvature or torsion of spacetime, but
only curvature or torsion of a connection. This becomes evident if we notice that many
different connections are allowed to exist in the very same spacetime. Of
course, when restricted to the specific case of general relativity, where the only
connection at work is the Levi--Civita connection, universality of gravitation allows it to be interpreted as part of the spacetime definition. However, in the presence of different connections with different curvatures and torsions, it seems far wiser and convenient to take
spacetime simply as a manifold, and connections (with their curvatures and torsions) as
additional structures. 

A spin connection $A_\mu$ is a connection of the form 
\be
A_\mu = \onehalf \, A^{ab}{}_\mu \, S_{ab},
\ee
with $S_{ab}=-S_{ba}$ Lorentz generators in a given representation. On the other hand, a tetrad field relates internal with external tensors. For example, if $V^a$ is a Lorentz vector,
\be
V^\rho = h_a{}^\rho \, V^a
\ee
will be a spacetime vector. However, in the specific case of connections, an additional {\it vacuum} term appears when transforming internal to external indices, and vice versa. In fact, a general linear connection $\Gamma^{\rho}{}_{\nu \mu}$ is related to the corresponding spin connection $A^{a}{}_{b \mu}$ through
\be
\Gamma^{\rho}{}_{\nu \mu} = h_{a}{}^{\rho} \partial_{\mu} h^{a}{}_{\nu} +
h_{a}{}^{\rho} A^{a}{}_{b \mu} h^{b}{}_{\nu}.
\label{geco}
\ee
The inverse relation is, consequently,
\be
A^{a}{}_{b \mu} =
h^{a}{}_{\nu} \partial_{\mu} h_{b}{}^{\nu} +
h^{a}{}_{\nu} \Gamma^{\nu}{}_{\rho \mu} h_{b}{}^{\rho}.
\label{gsc}
\ee
Equations (\ref{geco}) and (\ref{gsc}) are simply different ways of expressing the property
that the total --- that is, acting on both indices --- covariant derivative of the tetrad vanishes identically:
\be
\partial_{\mu} h^{a}{}_{\nu} - \Gamma^{\rho}{}_{\nu \mu} h^{a}{}_{\rho} +
A^{a}{}_{b \mu} h^{b}{}_{\nu} = 0.
\label{todete}
\ee

A connection $\Gamma^\rho{}_{\lambda\mu}$ is said to be metric compatible if
\be \label{fourm}
\partial_\lambda
g_{\mu\nu} - \Gamma^\rho{}_{\mu \lambda}g_{\rho\nu} -
\Gamma^\rho{}_{\nu \lambda} g_{\mu \rho} = 0.
\ee
From the tetrad point of view, by using Eqs.~(\ref{geco}) and (\ref{gsc}), this equation
can be rewritten in the form
\be
h_\mu (\eta_{ab}) - A^d{}_{a\mu} \, \eta_{db} - A^d{}_{b\mu} \, \eta_{ad} = 0,
\ee
where $h_\mu = h^a{}_\mu \partial_a$. Since $h_\mu (\eta_{ab}) = 0$, we obtain
\be
A_{ba\mu} = -\,  A_{ab\mu}.
\ee
The underlying content of the metric--preserving property is, therefore, that the spin
connection is Lorentzian. 

The curvature and the torsion of the connection $A^{a}{}_{b \mu}$ are defined
respectively by
\be
R^a{}_{b \nu \mu} = \partial_{\nu} A^{a}{}_{b \mu} -
\partial_{\mu} A^{a}{}_{b \nu} + A^a{}_{e \nu} A^e{}_{b \mu}
- A^a{}_{e \mu} A^e{}_{b \nu}
\ee
and
\be
T^a{}_{\nu \mu} = \partial_{\nu} h^{a}{}_{\mu} -
\partial_{\mu} h^{a}{}_{\nu} + A^a{}_{e \nu} h^e{}_{\mu}
- A^a{}_{e \mu} h^e{}_{\nu}.
\ee
Using the relation (\ref{gsc}), they can be expressed in a purely spacetime form:
\be
\label{sixbm}
R^\rho{}_{\lambda\nu\mu} \equiv h_a{}^\rho \, h^b{}_\lambda \, R^a{}_{b \nu \mu}
= \partial_\nu \Gamma^\rho{}_{\lambda \mu} -
\partial_\mu \Gamma^\rho{}_{\lambda \nu} +
\Gamma^\rho{}_{\eta \nu} \Gamma^\eta{}_{\lambda \mu} -
\Gamma^\rho{}_{\eta \mu} \Gamma^\eta{}_{\lambda \nu}
\ee
and
\be
T^\rho{}_{\nu \mu} \equiv h_a{}^\rho \, T^a{}_{\nu \mu} =
\Gamma^\rho{}_{\mu\nu}-\Gamma^\rho{}_{\nu\mu}.
\label{sixam}
\ee

The connection coefficients can be conveniently decomposed according to\footnote{The
magnitudes related with general relativity will be denoted with an over ``$\circ$''.}
\be
\Gamma^\rho{}_{\mu\nu} = {\stackrel{\circ}{\Gamma}}{}^{\rho}{}_{\mu \nu} +
K^\rho{}_{\mu\nu},
\label{prela0}
\ee
where
\be
{\stackrel{\circ}{\Gamma}}{}^{\sigma}{}_{\mu \nu} = {\textstyle
\frac{1}{2}} g^{\sigma \rho} \left( \partial_{\mu} g_{\rho \nu} +
\partial_{\nu} g_{\rho \mu} - \partial_{\rho} g_{\mu \nu} \right)
\label{lci}
\ee
is the torsionless Levi--Civita connection of general relativity, and
\be
K^\rho{}_{\mu\nu} = {\textstyle
\frac{1}{2}} \left(T_\nu{}^\rho{}_\mu + T_\mu{}^\rho{}_\nu -
T^\rho{}_{\mu\nu}\right)
\label{contor}
\ee
is the contortion tensor. In terms of the spin connection,  decomposition (\ref{prela0})
assumes the form
\be
A^c{}_{a\nu} = \Abol^c{}_{a\nu} + K^c{}_{a\nu},
\label{rela00}
\ee
where $\Abol^c{}_{a \nu}$ is the Ricci coefficient of rotation, the spin connection of
general relativity.

Now, since the spin connection is a tensor in the last index, we can write
\be
A^a{}_{bc} = A^a{}_{b \mu} \, h_c{}^\mu.
\ee
It can thus be easily verified that, in the anholonomic basis $h_a$, the curvature and
torsion components are given respectively by
\be \label{13bm}
R^a{}_{bcd} = h_c (A^a{}_{bd})  -
h_d (A^a{}_{bc}) + A^a{}_{ec} \, A^a{}_{bd}
- A^a{}_{ed} \, A^e{}_{bc} + f^e{}_{cd} \, A^a{}_{be}
\ee
and
\be
T^a{}_{bc} = - f^a{}_{bc} + A^a{}_{cb} - A^a{}_{bc}.
\label{13am}
\ee
Seen from this frame, therefore, torsion includes the anholonomy. Use of
(\ref{13am}) for three combinations of the indices gives
\begin{equation}%
A^{a}{}_{b c} = - \onehalf (f^{a}{}_{b c} + T^{a}{}_{b c} +
f_{b c}{}^{a} + T_{b c}{}^{a} + f_{c b}{}^{a} + T_{c b}{}^{a}).
\label{tobetaken2}
\end{equation}%
When torsion vanishes, as in general relativity, we obtain the usual expression of the
Ricci coefficient of rotation in terms of the anholonomy:
\begin{equation}%
\Abol^{a}{}_{b c} = -\,  \onehalf (f^{a}{}_{b c} +
f_{b c}{}^{a} + f_{c b}{}^{a}).
\label{tobetaken3}
\end{equation}%

We have now all tools necessary to study the possible roles played by torsion in gravitation. We begin by reviewing, in the next section, the basics of general relativity, Einstein's theory for gravitation. 

\section{General relativity}

Universality of both gravitational and inertial effects was one of the hints taken by Einstein in the way towards his theory. Another clue was the notion of field. This concept provides the best approach to interactions consistent with special relativity. All known forces are mediated by fields on spacetime. If  a field is to represent gravitation, it should, by the considerations above, be a universal field, equally felt by every particle. And, of all the fields present in a spacetime, the metric appears as the most fundamental. A gravitational field should, therefore,  be represented by a metric $g_{\mu \nu}$, with its absence being described by the flat Minkowski metric.

The relevant connection in  general relativity is the Levi--Civita connection~(\ref{lci}). It is the only Lorentz connection with vanishing torsion. It has, however, non-vanishing curvature, which is related to the presence of a gravitational field. In general relativity, therefore, torsion is chosen to vanish from the very beginning, and has no role in the description of the gravitational interaction. The minimal coupling prescription in this theory amounts to replace the usual ordinary derivative by a covariant derivative with the Levi--Civita connection. Acting on a spacetime vector field $V^\rho$, for example, it reads
\be
\nabol_\nu V^\rho = \partial_\nu V^\rho + \Gammabol^\rho{}_{\mu \nu} \, V^\mu.
\label{grcp}
\ee
Acting on a Lorentz vector $V^a$, it is
\be
\Dbol_\nu V^a = \partial_\nu V^a + \Abol^a{}_{b \nu} \, V^b.
\label{grcpbis}
\ee

The gravitational field  is described by the  Einstein--Hilbert Lagrangian
\begin{equation}
\Lbol = -\,  \frac{\sqrt{-g}}{2 k^2} \; \Rbol,
\label{eq:LagGR}
\end{equation}
where $g = \det(g_{\mu \nu})$, $k^2 = 8 \pi G/c^{4}$ and $\Rbol = g^{\lambda \rho} \, \Rbol^\nu{}_{\lambda \nu \rho}$ is the scalar curvature of the Levi--Civita connection. With  a matter (source) field represented by $\L_m$, the total Lagrangian is
\be
\L = \Lbol + \L_m.
\ee
Variation of the corresponding action integral with respect to the metric tensor $g^{\mu \nu}$ yields Einstein equation
\be
\Rbol_{\mu \nu} + \onehalf\,  g_{\mu \nu} \, \Rbol = k^2 \, \Theta_{\mu \nu},
\label{efe}
\ee
where $\Theta_{\mu \nu}$ is the symmetric source energy--momentum tensor.

On the other hand, the action integral of a particle of mass $m$ in a gravitational field is
\be
{\mathcal S} = -\, m c \int_a^b  ds,
\ee
with $ds = ({g_{\mu \nu} \, dx^\mu dx^\nu})^{1/2}$ the coordinate--independent spacetime line element. The corresponding equation of motion is consistent with the minimal coupling prescription (\ref{grcp}), and  is given by the geodesic equation
\be
\frac{d u^\rho}{d s} + {\Gammabol}{}^\rho{}_{\mu \nu} \; u^\mu \; u^\nu = 0.
\label{eq:geodesic}
\ee
This equation says simply that  the particle four-acceleration --- its left--hand side ---  vanishes. This property reveals the absence of the concept of gravitational {\em force}, a basic characteristic of the geometric description. In fact, instead of acting through a force, the presence of gravitation is supposed to produce a {\em curvature in spacetime}, the gravitational interaction being described by letting (spinless) particles to follow the geodesics of the  metric field. Notice that no other kind of spacetime deformation is supposed to exist. Torsion, which would be another natural spacetime deformation, is assumed to vanish from the start. This is the approach of general relativity, in which geometry replaces the concept of gravitational force, and the
trajectories are determined, not by force equations, but by geodesics. It is important to notice
that only an interaction presenting the property of universality can be described by a
geometrization of spacetime. It is also important to mention that, in the eventual lack of
universality,\footnote{Although the weak equivalence principle has passed all experimental tests and seems to be true at the classical level \cite{exp}, there are compelling evidences that it might not be valid at the quantum level \cite{vaxjo05}.} the general relativity description of gravitation would break down.

\section{Teleparallel gravity}

The first attempt to unify gravitation and electromagnetism was made by H.~Weyl in 1918 \cite{oft}. That proposal, though unsuccessful,  introduced for the first time the notions of {\em gauge transformations} and {\em gauge invariance}, and was the seed which has grown into today's gauge theories. Ten years after the introduction of torsion by E. Cartan in 1923 a second unification attempt was made by Einstein. It was based on the concept of distant  (or absolute) parallelism, also referred to as  teleparallelism. The crucial idea was the introduction of a tetrad field.  Since the specification of a  tetrad involves sixteen components, and the gravitational field, represented by the spacetime metric, requires only ten, the six additional degrees of freedom were  related by Einstein to  the electromagnetic field \cite{sauer}. This attempt  did not succeed either  but, like Weyl's,  introduced ideas that remain important to this day.

In fact, teleparallel gravity can be considered today a viable theory for gravitation \cite{review}. It can be interpreted as a gauge theory for the translation group:  the fundamental field   is the gauge potential $B_{\mu}$, a field assuming values in the Lie algebra of the translation group,
\be
B_{\mu} = B^{a}{}_{\mu} \, P_a,
\label{B}
\ee
where $P_{a} = \partial /\partial x^a$ are the translation generators. It  appears naturally as the nontrivial part of the tetrad field:
\be
h^{a}{}_{\mu} = \partial_{\mu}x^{a} + B^{a}{}_{\mu}.
\label{tetrada}
\ee
The fundamental connection of teleparallel gravity is the Weit\-zen\-b\"ock connection\footnote{It should be remarked that R. Weitzenb\"ock has never written such connection. Even though, this name has been commonly used to denote a particular Cartan connection with vanishing curvature.} which, in terms of the tetrad,  is written as\footnote{The magnitudes related with teleparallel gravity will be denoted with an over ``$\bullet$''.}
\be
\Gammaw^{\rho}{}_{\mu\nu} = h_a{}^\rho \, \partial_\nu h^a{}_\mu.
\label{wcon}
\ee
In contrast to Levi--Civita, it is a connection with non-vanishing torsion, but vanishing curvature. The Weitzenb\"ock and  Levi--Civita connections are 
related by
\be
\Gammaw^{\rho}{}_{\mu\nu} = \Gammabol^{\rho}{}_{\mu\nu} + \Kw^{\rho}{}_{\mu\nu},
\label{rela0}
\ee
where
\be
\Kw^{\rho}{}_{\mu\nu} = \onehalf \, (\Tw_\mu{}^\rho{}_\nu + \Tw_\nu{}^\rho{}_\mu - 
\Tw^{\rho}{}_{\mu\nu})
\ee
is the contortion tensor, with
\be
\Tw^{\rho}{}_{\mu\nu} = \Gammaw^{\rho}{}_{\nu\mu} - \Gammaw^{\rho}{}_{\mu\nu}
\ee
the torsion of the Weitzenb\"ock connection.

The coupling prescription in teleparallel gravity is obtained by requiring consistency with the general covariance principle, an active version of the strong equivalence principle \cite{weinberg}. It follows that it is actually equivalent to that of general relativity \cite{mospe}. Acting on a spacetime vector field $V^\rho$, for example, it is given by
\be
\nablaw_\nu V^\rho = \partial_\nu V^\rho + (\Gammaw^\rho{}_{\mu \nu} -
\Kw^\rho{}_{\mu \nu}) \, V^\mu.
\label{tgcp}
\ee
Since, as a consequence of  definition (\ref{wcon}), the Weitzenb\"ock spin connection vanishes identically,
\be
\Aw^{a}{}_{b \mu} =
h^{a}{}_{\nu} \partial_{\mu} h_{b}{}^{\nu} +
h^{a}{}_{\nu} \Gammaw^{\nu}{}_{\rho \mu} h_{b}{}^{\rho} = 0,
\ee
the corresponding covariant derivative of a Lorentz vector $V^a$ is \cite{equivcova}
\be
\Dw_\nu V^a = \partial_\nu V^a + (0 - \Kw^a{}_{b \nu}) \, V^b.
\label{tgcpbis}
\ee
Note that the covariant derivatives (\ref{tgcp}) and (\ref{tgcpbis}) are the Levi--Civita  derivatives rephrased in terms of the Weitzenb\"ock connection.

The Lagrangian of the teleparallel equivalent of general relativity is
\be
\Lw = \frac{h}{2 k^2} \; \left[\frac{1}{4} \;
\Tw^\rho{}_{\mu \nu} \; \Tw_\rho{}^{\mu \nu} + \frac{1}{2} \;
\Tw^\rho{}_{\mu \nu} \; \Tw^{\nu \mu} {}_\rho - \Tw_{\rho \mu}{}^{\rho}
\; \Tw^{\nu \mu}{}_\nu \right],
\label{lagr3}
\ee
where $h = \det (h^a{}_\mu)$. The first term corresponds to the usual Lagrangian of gauge theories. However, owing to the presence of  tetrad fields, algebra and spacetime indices can here be
changed into each other, and new contractions turn out to be possible. It
is exactly this possibility that gives rise to the other two terms. If we define the tensor
\begin{equation}
\Sw^{\rho\mu\nu} = -\,  \Sw^{\rho\nu\mu} =
\left[ \Kw^{\mu\nu\rho} - g^{\rho\nu}\,\Tw^{\sigma\mu}{}_{\sigma}
+ g^{\rho\mu}\,\Tw^{\sigma\nu}{}_{\sigma} \right],
\label{S}
\end{equation}
the teleparallel Lagrangian (\ref{lagr3}) can be rewritten as \cite{maluf94}
\begin{equation}
\Lw =
\frac{h}{4 k^2} \; \Sw^{\rho\mu\nu} \, \Tw_{\rho\mu\nu}.
\label{gala}
\end{equation}
Using relation (\ref{rela0}), it is easy to show that
\begin{equation}
\Lw = \Lbol - \partial_\mu \left(2 \, h \, k^{-2} \,
\Tw^{\nu \mu}{}_\nu \right),
\end{equation}
where $\Lbol$ is the Einstein--Hilbert Lagrangian (\ref{eq:LagGR}), and where we have used $h = \sqrt{-g}$. Up to a divergence, therefore, the teleparallel Lagrangian is equivalent to the Lagrangian of general relativity.

Let us consider now
\begin{equation}
{\mathcal L} = \Lw + {\mathcal L}_m,
\end{equation}
with ${\mathcal L}_m$ the Lagrangian of a general matter field. Variation of the corresponding action integral with respect to the gauge field $B^a{}_\rho$ leads to the teleparallel version of the gravitational field equation
\be
\partial_\sigma(h\, \Sw_\mu{}^{\rho \sigma}) -
k^2 \, (h\, \tw_{\mu}{}^{\rho}) = k^2 \, (h\,  {\Theta}_{\mu}{}^{\rho}),
\label{tfe1}
\ee
where $\tw_{\mu}{}^{\rho}$ represents the energy--momentum pseudotensor of the gravitational field~\cite{gemt}, and ${\Theta}_{\mu}{}^{\rho}$ is the symmetric source energy--momentum tensor. Using relation (\ref{rela0}), the left-hand side of the field equation (\ref{tfe1}) can be shown to be
\begin{equation}
\partial_\sigma(h \Sw_\mu{}^{\rho \sigma}) -
k^2 \, (h \tw_{\mu}{}^{\rho}) =
h \left({\stackrel{\circ}{R}}_\mu{}^{\rho} -
\onehalf \, \delta_\mu{}^{\rho} \;
{\stackrel{\circ}{R}} \right).
\label{ident}
\end{equation}
This means that, as expected due to the equivalence between the cor\-re\-sponding Lagrangians, the teleparallel field equation (\ref{tfe1}) is equivalent to Einstein field equation (\ref{efe}). Observe that the symmetric energy--momentum tensor appears as the source in both theories: as the source of curvature in general relativity, and as the source of torsion in teleparallel gravity.

In teleparallel gravity, the action describing a particle of mass $m$ in a gravitational field $B^a{}_\mu$ is given by
\be
{\mathcal S} = -\, m \, c \int_{a}^{b} \left[u_a \, dx^a +
B^{a}{}_{\mu} \, u_{a} \, dx^{\mu} \right],
\label{acaopuni0}
\ee
where $u^a = h^a{}_\mu \, u^\mu$ is the anholonomic particle four--velocity. The first term represents the action of a free particle, and the second the coupling of the particle's mass with the gravitational field. The corresponding equation of motion is consistent with the coupling prescription (\ref{tgcp}), and is given by the {\it force equation}~\cite{paper1}
\be
\frac{d u_\mu}{d s} -
\Gammaw^\theta{}_{\mu \nu} \; u_\theta \; u^\nu =
\Tw^\theta{}_{\mu \nu} \; u_\theta \, u^\nu,
\label{forceq}
\ee
with torsion playing the role of {\it gravitational force}. It is similar to the Lorentz force e\-qua\-tion of electrodynamics, a property related to the fact that teleparallel gravity is, like Maxwell's theory, a gauge theory. By using relation (\ref{rela0}), the force equation (\ref{forceq}) can be rewritten in terms of the Levi--Civita connection, in which case it reduces to the geodesic 
equation (\ref{eq:geodesic}). The force equation (\ref{forceq}) of teleparallel gravity and the geodesic equation (\ref{eq:geodesic}) of general relativity describe, therefore,  the same physical trajectory. This means that the gravitational interaction has two {\em equivalent} descriptions: one in terms of curvature, and another in terms of torsion. Although equivalent, however, there are conceptual differences between these two descriptions. In general relativity, a theory based on the weak equivalence principle, curvature is used to {\it geometrize} the gravitational interaction. In teleparallel gravity, on the other hand, torsion accounts for gravitation, not by geometrizing the interaction, but by acting as a {\it force}. Like Maxwell theory, there are no geodesics in teleparallel gravity, only force equations.

An important property of teleparallel gravity is that, due to its gauge structure, it does not require the weak equivalence principle to describe the gravitational interaction \cite{wep}.
To understand this point, let us consider a particle with inertial mass $m_i$ and gravitational mass $m_g$. In this case, the teleparallel action is written as
\be
{\mathcal S} = - \, m_i \, c \int_{a}^{b} \left[u_a \, d{x}^a +
\frac{m_g}{m_i} \, B^{a}{}_{\mu} \, u_{a} \, dx^{\mu} \right].
\label{acaopuni}
\ee
Observe the similarity with the action
\be
{\mathcal S} = - \, m_i \, c \int_{a}^{b} \left[u_a \, d{x}^a +
\frac{q}{m_i} \, A_{a} \, dx^{a} \right],
\label{acaopuni2}
\ee
which describes a particle with mass $m_i$ and electric charge $q$ in an electromagnetic field $A_a$. We see from these expressions that the electric charge $q$ plays the same role as the gravitational mass $m_g$. Variation of (\ref{acaopuni}) yields
\be
P^\rho{}_\mu
\left(\partial_\rho {x}^a + \frac{m_g}{m_i} \, {B}^a{}_\rho \right) \frac{d u_a}{d s} =
\frac{m_g}{m_i} \; \Tw^a{}_{\mu \rho} \; u_a \, u^\rho,
\label{eqmot3bis}
\ee
with $P^\rho{}_\mu = \delta^\rho_\mu - u^\rho \, u_\mu$ a projection tensor. This is the equation of motion for particles with $m_g \neq m_i$ in the presence of gravitation. For $m_i = m_g$, it reduces to the teleparallel force equation (\ref{forceq}), which in turn is equivalent to the geodesic equation (\ref{eq:geodesic}) of general relativity. It is, however, impossible to get this kind of  equation in the context of general relativity --- which is not valid if there is no universality. In other words, whereas the geometrical description of general relativity breaks down, the gauge description of teleparallel gravity stands up in the lack of universality.\footnote{Differently from general relativity, both teleparallel and Newtonian gravities are able to manage without the weak equivalence principle. Furthermore, since these two theories describe the gravitational interaction by a force equation, the Newtonian limit is found to follow much more naturally from teleparallel gravity than from general relativity.} This is a very important issue because, even though the equivalence principle has got through many experimental tests, there are many controversies related with its validity \cite{synge}, mainly at the quantum level \cite{quantu}.

One may wonder why gravitation has two equivalent descriptions. This duplicity is 
related precisely to that peculiar property of gravitation, universality. As remarked above, gravitation can be described in terms of a gauge theory --- just teleparallel gravity. Universality of free fall, on the other hand, makes it possible a second, geometrized description, based on the weak equivalence principle --- just general relativity. As the sole universal interaction, it is the only one to allow also a geometrical interpretation, and hence two alternative descriptions. From this point of view, curvature and torsion are simply alternative ways of describing the gravitational field \cite{aap}, and consequently related to the same degrees of freedom. The gravitational interaction can thus be described {\em alternatively} in terms of curvature, as is usually done in general relativity, or in terms of torsion, in which case we have the so-called teleparallel gravity. Accordingly, we can say that, from the point of view of teleparallel gravity, Einstein was right when he did not include torsion in general relativity.

\section{Einstein--Cartan theory}

The main idea behind the Einstein--Cartan construction \cite{ect} is that, at the microscopic level, matter is represented by elementary particles, which in turn are characterized by mass (that is, energy and momentum) and spin. If one adopts the same {\it geometrical spirit of general relativity}, not only mass but also spin should be source of gravitation at this level. According to this scheme, energy--momentum should appear as source of curvature, whereas spin should appear as source of torsion. 

The relevant connection of this theory is a general Cartan connection $\Gamma^\rho{}_{\mu \nu}$, presenting both curvature and torsion. Similarly to general relativity, the Lagrangian of the gravitational field in Einstein--Cartan theory is
\begin{equation}
{\mathcal L}_{EC} = -\,  \frac{\sqrt{-g}}{2 k^2} \; R.
\end{equation}
Observe that, although it formally coincides with the Einstein--Hilbert Lagrangian, the scalar curvature refers now to the general Cartan connection. Considering then the Lagrangian
\be
{\mathcal L} = {\mathcal L}_{EC} + {\mathcal L}_m,
\ee
with ${\mathcal L}_m$ the Lagrangian of a general matter field, the ensuing field equations are obtained through variations with respect to the metric $g^{\mu \nu}$ and to the torsion $T_\rho{}^{\mu \nu}$. The result are
\be
R_{\mu \nu} - \onehalf\,  g_{\mu \nu} R = k^2 \, \theta_{\mu \nu}
\ee
and
\be
T^\rho{}_{\mu \nu} = k^2 \left( S^\rho{}_{\mu \nu} +
\onehalf\,  \delta^\rho{}_\mu \, S^\alpha{}_{\alpha \nu} -
\onehalf\, \delta^\rho{}_\nu \, S^\alpha{}_{\alpha \mu} \right),
\ee
where $\theta_{\mu \nu}$ is the {\it canonical} energy--momentum tensor of the source, which is related to the symmetric energy--momentum tensor $\Theta_{\mu \nu}$ through the Belinfante--Rosenfeld procedure \cite{br}, and $S^\rho{}_{\mu \nu}$ is the spin tensor.

An emblematic property of Einstein--Cartan theory is that the field equation for torsion is an algebraic equation, and consequently torsion is a non-propagating field. In spite of this peculiarity, this theory can be considered as a paradigm of more general gravitational models --- like gauge theories for the Poincar\'e \cite{kibble} and the affine groups \cite{hcmn} --- in the sense that all these models presuppose new physics associated to torsion. In other words, curvature and torsion in these theories represent independent gravitational degrees of freedom. We can then say that, from the point of view of the Einstein--Cartan theory, as well as of the more general gauge theories for gravitation, Einstein made a mistake by neglecting torsion.

The coupling prescription in Einstein--Cartan theory is usually assumed to be given by the covariant derivative in terms of the connection $\Gamma^\rho{}_{\mu \nu}$. Acting on a spacetime vector field $V^\rho$, for example, it reads
\be
\nabla_\nu V^\rho = \partial_\nu V^\rho + \Gamma^\rho{}_{\mu \nu} \, V^\mu,
\label{eccp}
\ee
whereas acting on a Lorentz vector $V^a$, it is
\be
{\mathcal D}_\nu V^a = \partial_\nu V^a + A^a{}_{b \nu} \, V^b.
\ee
Now, in this theory, the equation of motion of particles is usually obtained by considering the generalized matter energy--momentum covariant conservation law, integrating over a space-like section of the world tube of the particle, and expanding the gravitational field in power series~\cite{Pap51}. For spinning particles, in addition to the usual Papapetrou coupling between particle's spin with the Riemann tensor, there will appear in the equation of motion a coupling between spin and torsion. For spinless particles, it reduces to the geodesic equation (\ref{eq:geodesic}). Differently from general relativity and teleparallel gravity, therefore, where the equations of motion of spinless particles are obtained by replacing the ordinary differential by the corresponding covariant differential, the equation of motion for such particles in Einstein--Cartan theory does not follow from the minimal coupling prescription. To a certain extent, and considering the crucial role played by the minimal coupling prescription in the description of the fundamental interactions, this point can be considered a drawback of the Einstein--Cartan model. Furthermore, the coupling prescription (\ref{eccp}) presents some additional problems: it is not consistent \cite{ap0} with the general covariance principle \cite{weinberg} --- an active version of the usual (passive) strong equivalence principle --- and when applied to describe the interaction of the electromagnetic field with gravitation, it violates the U(1) gauge invariance of Maxwell's theory.

\section{New general relativity}

As already remarked, the teleparallel structure was used by Einstein in his unsuccessful attempt to unify gravitation and electromagnetism. In the sixties, M{\o}ller~\cite{moller} revived the idea of teleparallelism, but then with the sole purpose of describing gravitation. Afterwards, Pellegrini \& Plebanski~\cite{pelle} found a Lagrangian formulation for teleparallel gravity, a problem that was reconsidered later by M{\o}ller~\cite{moller2}. In 1967, Hayashi \& Nakano~\cite{haya} formulated a gauge model for the translation group. A few years later, Hayashi~\cite{hay77} pointed out the connection between that theory and teleparallelism, and an attempt to unify these two developments was made by Hayashi \& Shirafuji~\cite{hs79} in 1979. In this approach, general relativity --- or its teleparallel equivalent --- is supplemented with a generalized teleparallel gravity, a theory that involves only torsion, and presents three free parameters, to be determined by experiment.

Like in the teleparallel equivalent of general relativity, the relevant connection of new general relativity is the Weitzenb\"ock connection (\ref{wcon}). The coupling prescription, however, is assumed to be given by a covariant derivative in terms of the Weitzenb\"ock connection:
\be
\dw_\nu V^\rho = \partial_\nu V^\rho + \Gammaw^\rho{}_{\mu \nu} \, V^\mu.
\label{ngrcp}
\ee
Since the Weitzenb\"ock spin connection vanishes identically, $\Aw^a{}_{b \nu} = 0$, the corresponding covariant derivative of a Lorentz vector $V^a$ will coincide with an ordinary derivative \cite{hs79}:
\be
\dw_\nu V^\rho = \partial_\nu V^\rho.
\ee
Considering that, like in Einstein--Cartan theory, the equation of motion of spinless particles in new general relativity is the geodesic equation (\ref{eq:geodesic}), here also there is an inconsistency between the coupling prescription and the equation of motion of spinless particles.

The Lagrangian of the gravitational field in new general relativity has the form
\be
{\mathcal L}_{ngr} = \frac{h}{2 k^2} \; \left[a_1 \,
\Tw^\rho{}_{\mu \nu} \; \Tw_\rho{}^{\mu \nu} + a_2 \,
\Tw^\rho{}_{\mu \nu} \; \Tw^{\nu \mu} {}_\rho + a_3 \, \Tw_{\rho \mu}{}^{\rho}
\; \Tw^{\nu \mu}{}_\nu \right],
\label{ngrlag0}
\ee
with $a_1, a_2, a_3$ arbitrary coefficients. Now, as is well known, torsion can be decomposed in irreducible components under the global Lorentz group \cite{hb73}:
\be
\Tw_{\lambda \mu \nu} = \textstyle{\frac{2}{3}} \left(t_{\lambda \mu \nu} -
t_{\lambda \nu \mu} \right) + \frac{1}{3} \left(g_{\lambda \mu} v_\nu -
g_{\lambda \nu} v_\mu \right) + \epsilon_{\lambda \mu \nu \rho} \, a^\rho.
\label{deco}
\ee
In this expression, $v_\mu$ and $a^\rho$ are the vector and axial parts of torsion,
defined respectively by
\be
v_{\mu} =  \Tw^{\nu}{}_{\nu \mu} \quad \mbox{and} \quad
a^{\mu} = \textstyle{\frac{1}{6}} \epsilon^{\mu\nu\rho\sigma} \, \Tw_{\nu\rho\sigma},
\label{pt3}
\ee
whereas $t_{\lambda \mu \nu}$ is the purely tensor part, given by
\be
t_{\lambda \mu \nu} = \textstyle{\frac{1}{2}} \left(\Tw_{\lambda \mu \nu} +
\Tw_{\mu\lambda \nu} \right) + \frac{1}{6} \left(g_{\nu \lambda} v_\mu +
g_{\nu \mu} v_\lambda \right) - \frac{1}{3} g_{\lambda \mu} \, v_\nu.
\label{pt1}
\ee
In terms of these components, the above Lagrangian reads
\be
{\mathcal L}_{ngr} = \frac{h}{2 k^2} \; \left[b_1 \,
t^\rho{}_{\mu \nu} \, t_\rho{}^{\mu \nu} + b_2 \,
v^\mu \, v_\mu + b_3 \, a^\mu \, a_\mu \right],
\label{ngrlag1}
\ee
with $b_1, b_2, b_3$ new arbitrary coefficients. Considering then the identity
\be
\textstyle \frac{2}{3} \, t^\rho{}_{\mu \nu} \, t_\rho{}^{\mu \nu} +
\frac{2}{3} \, v^\mu \, v_\mu - \frac{3}{2} \, a^\mu \, a_\mu = \Rbol,
\label{iden3}
\ee
it can be rewritten in the form
\be
{\mathcal L}_{ngr} = \frac{h}{2 k^2} \; \left[ \Rbol + c_1 \,
t^\rho{}_{\mu \nu} \, t_\rho{}^{\mu \nu} + c_2 \,
v^\mu \, v_\mu + c_3 \, a^\mu \, a_\mu \right],
\label{ngrlag2}
\ee
with
\be
\textstyle c_1 = b_1 - \frac{2}{3}, \quad c_2 = b_2 -
\frac{2}{3}, \quad c_3 = b_3 + \frac{3}{2}.
\ee
According to this theory, therefore, torsion is assumed to produce deviations from the predictions of general relativity --- or equivalently, from the predictions of the teleparallel equivalent of general relativity. This means that, similarly to Einstein--Cartan theory, torsion represents additional degrees of freedom of gravity. Also from the point of view of new general relativity, therefore, Einstein made a mistake by neglecting  torsion.

It should be remarked that solar system experiments restrict severely the existence of non-vanishing $c_1$ and $c_2$. Furthermore, as already shown in the literature \cite{op}, the Schwarzschild solution exists only for the case with $c_1 = c_2 = c_3 = 0$. In principle, therefore, we can say that new general relativity lacks experimental support. Anyway, there has been recently a proposal to look for some eventual effects produced by a non--vanishing $c_3$ using the Gravity Probe B data \cite{mao}. The idea behind such proposal lies on the fact that the axial torsion $a^\mu$, which represents the gravitomagnetic component of the gravitational field \cite{apz}, is responsible for producing the Lense--Thirring effect, which is one of the effects Gravity Probe B was intended to detect.

\section{Final remarks}

In general relativity, curvature represents the gravitational field. In teleparallel gravity, it is torsion that represents the gravitational field. In spite of this fundamental difference, the two theories are found to yield equivalent descriptions of the gravitational interaction. An immediate implication of this equivalence is that curvature and torsion are simply alternative ways of describing the gravitational field, and are consequently related to the same degrees of freedom. This is corroborated by the fact that the symmetric matter energy-momentum tensor appears as source in both theories: as source of curvature in general relativity, as source of torsion in teleparallel gravity.

Now, more general gravity theories, like Einstein-Cartan, gauge theories for the Poincar\'e and more general groups, as well as new general relativity, consider curvature and torsion as representing independent degrees of freedom. In these theories, therefore, torsion describes additional degrees of freedom and, in consequence, new physical phenomena should be expected.
These differences give rise to a conceptual question concerning the actual role played by torsion. The two points of view are physically conflictive: if one is correct, the other is necessarily wrong. Which of them is right? In principle, experience should give the answer, but this is not so simple --- there seems to be no  model--independent way to look for torsion. For example, due to the Einstein--Cartan theory, there is a widespread belief that  torsion has an intimate association with spin, and is consequently important only at the microscopic level. Most searches rely on this point of view \cite{exptor}, though a recent proposal \cite{mao} looks for effects as predicted by new general relativity.

It should be remarked that, due to the weakness of the gravitational interaction, there are no available data on the gravitational coupling of the fundamental particles. Concerning macroscopic physics, no one has ever reported new gravitational phenomena near a neutron star, for example, where the effects of torsion would be relevant according to Einstein--Cartan theory. Actually, there are no experimental signs of torsion in the sense predicted by Einstein--Cartan, gauge theories for the Poincar\'e and more general groups, and new general relativity. On the other hand, according to teleparallel gravity, torsion has already been detected: it is responsible for all known gravitational effects, including the physics of the solar system, which can be reinterpreted in terms of a force equation, with torsion playing the role of gravitational force. We could then say that the existing experimental data favor the teleparallel point of view, and consequently general relativity.

From the conceptual point of view, all alternative models --- Einstein--Cartan, gauge theories for Poincar\'e and more general groups, as well as new general relativity --- present consistency problems. For example, even though the coupling prescription of these models can comply with the {\it passive} strong equivalence principle, they are not consistent \cite{ap0} with the {\em active} version of the strong equivalence principle, also known as the general covariance principle \cite{weinberg}. Another relevant problem is that, when used to describe the interaction of the electromagnetic field with gravitation, the coupling prescriptions of these models violate the U(1) gauge invariance of Maxwell's theory. This problem is usually circumvented by {\it postulating} that the electromagnetic field does not couple to torsion \cite{postulate}. This ostrich-like ``solution'' is, however, far from reasonable. On the other hand, the teleparallel interpretation for torsion presents several conceptual advantages in relation to the other theories: it is consistent with both active and passive versions of the strong equivalence principle \cite{mospe} and describes the interplay of electromagnetic and gravitational fields without violating electromagnetic gauge invariance \cite{vector}. In spite of the conceptual soundness of the teleparallel approach, we prefer to say once more  that a definitive answer can only be achieved by experiments.

\section*{Acknowledgments}
The authors would like to thank FAPESP, CNPq and CAPES for partial 
financial support.



\begin{thebibliography}{99}

\bibitem{mtw}
C. W. Misner, K. S. Thorne and J. A. Wheeler, {\it Gravitation} (Freeman, New York,
1973).

\bibitem{moriond}
A simplified analysis of these questions has appeared in J. G. Pereira, {\it In Search of the Spacetime Torsion}. Talk presented at the ``Rencontres de Moriond on Gravitational Waves and Experimental Gravity'', La Thuile, Italy, March 11-18, 2007
[gr-qc/0704.1141].

\bibitem{livro}
R. Aldrovandi and J. G. Pereira, {\it An Introduction to Geometrical Physics}
(World Scientific, Singapore, 1995).

\bibitem{ABP03}
R. Aldrovandi, P. B. Barros  and  J. G. Pereira,  {\em Gen. Rel. Grav.} {\bf 35}, 991 (2003) [gr-qc/0301077].

\bibitem{koba}
S. Kobayashi and K. Nomizu, {\it Foundations of Differential Geometry}
(Interscience, New York, 1963).

\bibitem{exp}
C. M. Will, {\it Living Rev. Rel.} {\bf 9}, 3 (2006) [gr-qc/0510072].

\bibitem{vaxjo05}
R. Aldrovandi, J. G. Pereira and K. H. Vu, {\it Gravity and the Quantum: Are They Reconcilable?} Proceedings of the conference ``Quantum Theory: Reconsideration of Foundations-3'' (AIP Conference Proceedings, New York, 2006) Vol. 810, page 217 [gr-qc/0509051].

\bibitem{oft}
On the birth and evolution of gauge theories, see  L. O'Raifeartaigh, {\em The Dawning of Gauge Theory} (Princeton University Press, Princeton, 1998); see and also L. O'Raifeartaigh and N. Straumann, {\it Rev. Mod. Phys.} {\bf 72}, 1 (2000).

\bibitem{sauer}
A description of the teleparallel--based Einstein's unification theory can be found in T. Sauer, {\it Field equations in teleparallel spacetime: Einstein's `Fern\-paral\-le\-lis\-mus' approach towards unified field theory}, Einstein's Papers Project [{physics/0405142}].

\bibitem{review}
H. I. Arcos and J. G. Pereira, {\it Int. J. Mod. Phys.}  
{\bf D13}, 2193 (2004) [gr-qc/0501017].

\bibitem{weinberg}
S. Weinberg, {\it Gravitation and Cosmology} (Wiley, New York, 1972).

\bibitem{mospe}
R. A. Mosna and J. G. Pereira, {\it Gen. Rel. Grav.} {\bf 36}, 2525 (2004) [gr-qc/0312093].

\bibitem{equivcova}
V. C. de Andrade, L. C. T. Guillen and J. G. Pereira, 
{\it Phys. Rev.} {\bf D64}, 027502 (2001) [gr-qc/0104102].

\bibitem{maluf94}
J. W. Maluf, {\it J. Math. Phys.} {\bf 35}, 335 (1994).

\bibitem{gemt}
V. C. de Andrade, L. C. T. Guillen and J. G. Pereira, {\it Phys. Rev. Lett.} {\bf 84}, 4533 (2000) [gr-qc/0003100].

\bibitem{paper1}
V. C. de Andrade and J. G. Pereira, {\it Phys. Rev.}  {\bf D56}, 4689 (1997) 
[gr-qc/9703059].

\bibitem{wep}
R. Aldrovandi, J. G. Pereira and K. H. Vu, {\it Gen. Rel. Grav.} {\bf 36}, 
101 (2004) [gr-qc/0304106].

\bibitem{synge}
For a critical discussion of the equivalence principle, see the preface of J. L. Synge,
{\it Relativity: The General Theory} (North-Holland, Amsterdam, 1960). See also
T. Damour, {\it Comptes Rendus de l'Acad\'emie des Sciences (Serie IV)} {\bf 2}, 1249 (2001)
[{gr-qc/0109063}]; R. Aldrovandi, P. B. Barros and J. G. Pereira, 
{\em Found. Phys.} {\bf 33}, 545 (2003) [gr-qc/0212034]. 

\bibitem{quantu}
M. P. Haugan and C. L\"amerzahl, {\it Lect. Notes Phys.} {\bf 562}, 195 (2001);
C. L\"ammerzahl, {\it Gen. Rel. Grav.} {\bf 28}, 1043 (1996); C. L\"ammerzahl,
{\it Acta Phys. Pol.} {\bf 29}, 1057 (1998); C. L\"ammerzahl,
{\it Class. Quant. Grav.} {\bf 15}, 13 (1998).

\bibitem{aap}
H. I. Arcos, V. C. de Andrade and  J. G. Pereira, {\it Int. J. Mod. Phys.} 
 {\bf D13}, 807 (2004) [gr-qc/0403074].

\bibitem{ect}
For a textbook reference on the Einstein--Cartan theory, see
V. de Sabbata and M. Gasperini, {\em Introduction to Gravitation}
(World Scientific, Singapore, 1985).

\bibitem{br}
F. J. Belinfante, {\em Physica} {\bf 6}, 687 (1939); 
L. Rosenfeld, {\em M\'em. Acad. Roy. Belg. Sci.} {\bf 18}, 1 (1940).

\bibitem{kibble}
T. W. B. Kibble, {\it J. Math. Phys.} {\bf 2}, 212 (1961); F. Gronwald and F. W. Hehl, in {\it Proceedings of the 14th School of Cosmology and Gravitation}, Erice, Italy, ed. by P. G. Bergmann, V. de Sabbata and H.-J Treder (World Scientific, Singapore, 1996); 
R. Aldrovandi and J.G. Pereira, {\em  Phys. Rev.} {\bf D33}, 2788 (1986).

\bibitem{hcmn}
F. W. Hehl, J. D. McCrea, E. W. Mielke and Y. Ne'emann, {\it Phys. Rep.} {\bf 258}, 1
(1995).

\bibitem{Pap51}
A. Papapetrou, {\em Proc. R. Soc. London}, {\bf A209}, 248 (1951).

\bibitem{ap0}
H. I. Arcos and J. G. Pereira, {\it Class. Quant. Grav.} {\bf 21}, 5193 (2004)
[gr-qc/0408096].

\bibitem{moller}
C. M{\o}ller, {\it K. Dan. Vidensk. Selsk. Mat. Fys. Skr.} {\bf 1}, No.
10 (1961).

\bibitem{pelle}
C. Pellegrini and J. Plebanski, {\it K. Dan. Vidensk. Selsk. Mat. Fys. Skr.} {\bf 2}, No. 2 (1962).

\bibitem{moller2}
C. M{\o}ller, {\it K. Dan. Vidensk. Selsk. Mat. Fys. Skr.} {\bf 89}, No.
13 (1978).

\bibitem{haya}
K. Hayashi and T. Nakano, {\it Prog. Theor. Phys.} {\bf 38}, 491 (1967).

\bibitem{hay77}
K. Hayashi, {\it Phys. Lett.} {\bf B69}, 441 (1977).

\bibitem{hs79}
K. Hayashi and T. Shirafuji, {\it Phys. Rev.}  {\bf D19}, 3524 (1979).

\bibitem{hb73}
K. Hayashi and A. Bregman, {\it Ann. Phys. (NY)} {\bf 75}, 562 (1973).

\bibitem{op}
Yu. N. Obukhov and J. G. Pereira, {\it Phys. Rev.}  {\bf D67}, 044016 (2003) 
[gr-qc/0212080].

\bibitem{mao}
Y. Mao, M. Tegmark, A. Guth and S. Cabi, {\it Phys. Rev. D} (2007), in press [gr-qc/0608121].

\bibitem{apz}
J. G. Pereira T. Vargas and C. M. Zhang, {\it Class. Quant. Grav.} {\bf 18}, 833 (2001)
[gr-qc/0102070].

\bibitem{exptor}
C. L\"ammerzahl, {\it Phys. Lett.} {\bf A228}, 223 (1997);
R. T. Hammond, {\it Rep. Prog. Phys.} {\bf 65}, 599 (2002).

\bibitem{postulate}
For a discussion, see V. de Sabbata and C. Sivaram, {\it Spin and Torsion in 
Gravitation} (World Scientific, Singapore, 1994).

\bibitem{vector}
V. C. de Andrade and J. G. Pereira, {\it Int. J. Mod. Phys.}  {\bf D8}, 141 
(1999) [gr-qc/9708051].

\end{thebibliography}
\end{document}